\documentclass[%
 pra,twocolumn,
superscriptaddress,
 amsmath,amssymb,
 aps,
floatfix,
]{revtex4-1}

\usepackage{graphicx}
\usepackage{dcolumn}
\usepackage{bm}
\usepackage{color}
\usepackage{amsfonts}
\usepackage{graphicx}
\usepackage{amsmath}
\usepackage{times}
\usepackage{ulem}

\newcommand{\Yb}{^{171}{\rm Yb}^+}

\newcommand{\ket}[1]{\left|#1\right\rangle}
\newcommand{\bra}[1]{\left \langle #1\right |}
\newcommand{\colvec}[1]{|#1\rangle\!\rangle}
\newcommand{\rowvec}[1]{\langle\!\langle #1 |}

\newcommand{\hsmatele}[3]{\langle\!\langle #1 | #2 | #3 \rangle\!\rangle}
\newcommand{\set}[1]{\left\{ #1 \right\}}
\newcommand{\beginsupplement}{%
        \setcounter{table}{0}
        \renewcommand{\thetable}{S\arabic{table}}%
        \setcounter{figure}{0}
        \renewcommand{\thefigure}{S\arabic{figure}}%
     }
		
\begin{document}
	
	
	\title{Error-Mitigated Quantum Gates Exceeding Physical Fidelities in a Trapped-Ion System}
	\author{Shuaining Zhang,$^{1}$ Yao Lu,$^{1}$ Kuan Zhang,$^{2,1}$ Wentao Chen,$^{1}$ Ying Li,$^{3\ast}$ Jing-Ning Zhang,$^{1\ast}$ Kihwan Kim$^{\ast}$}
	\affiliation{
		\normalsize{Center for Quantum Information, Institute for Interdisciplinary Information Sciences, Tsinghua University, Beijing 10084, P. R. China}\\
		\normalsize{$^{2}$MOE Key Laboratory of Fundamental Physical Quantities Measurements, Hubei Key Laboratory of Gravitation and Quantum Physics, PGMF and School of Physics, Huazhong University of Science and Technology, Wuhan 430074, China}\\
		\normalsize{$^{3}$Graduate School of China Academy of Engineering Physics, Beijing 100193, China}\\
	}

	\date{\today}
	
	\begin{abstract}
		Various quantum applications can be reduced to estimating expectation values, which are inevitably deviated by operational and environmental errors. Although errors can be tackled by quantum error correction, the overheads are far from being affordable for near-term technologies. To alleviate the detrimental effects of errors, quantum error mitigation techniques have been proposed, which require no additional qubit resources. Here, we benchmark the performance of a quantum error mitigation technique based on probabilistic error cancellation in a trapped-ion system. Our results clearly show that effective gate fidelities exceed physical fidelities, i.e. we surpass the break-even point of eliminating gate errors, by programming quantum circuits. The error rates are effectively reduced from $(1.10\pm 0.12)\times10^{-3}$ to $(1.44\pm 5.28)\times10^{-5}$ and from $(0.99\pm 0.06)\times10^{-2}$ to $(0.96\pm 0.10)\times10^{-3}$ for single- and two-qubit gates, respectively. Our demonstration opens up the possibility of implementing high-fidelity computations on a near-term noisy quantum device.
	\end{abstract}
	
	\pacs{Valid PACS appear here}
	\maketitle

	Quantum computers~\cite{feynman1982simulating} can extend classical computational reach in diverse research fields, including quantum chemistry, material science, and even machine learning. Based on various technological advances so far, such nontrival quantum applications have been pursued with currently available devices mainly through quantum-classical hybrid schemes~\cite{mcclean2016theory,peruzzo2014variational}. The schemes combine the advantages of classical and quantum computation, where quantum processors are used to estimate expectation values of physical observables on certain states for classical feedback. The hybrid schemes can be applied to estimate the ground state energies of molecules \cite{peruzzo2014variational,shen2017quantum,kandala2017hardware}, to simulate quantum models in materials\cite{PhysRevX.6.031045} and high-energy physics \cite{kokail2018self} and to find approximate solutions of optimization problems \cite{farhi2014quantum}.  Although it is anticipated that around a hundred well-behaved qubits are required for such schemes to outperform current classical counterparts in quantum chemistry~\cite{wecker2014gate,reiher2017elucidating,Simon2019review}, the advantages are only possible with accurate quantum processors. However, output results of the quantum devices are inevitably deviated because of errors originated from both environmental fluctuations and operational imperfections. Therefore, techniques for improving the accuracy of noisy quantum processors are of great importance.
	
		\begin{figure*}[ht]
		\includegraphics[width = \textwidth]{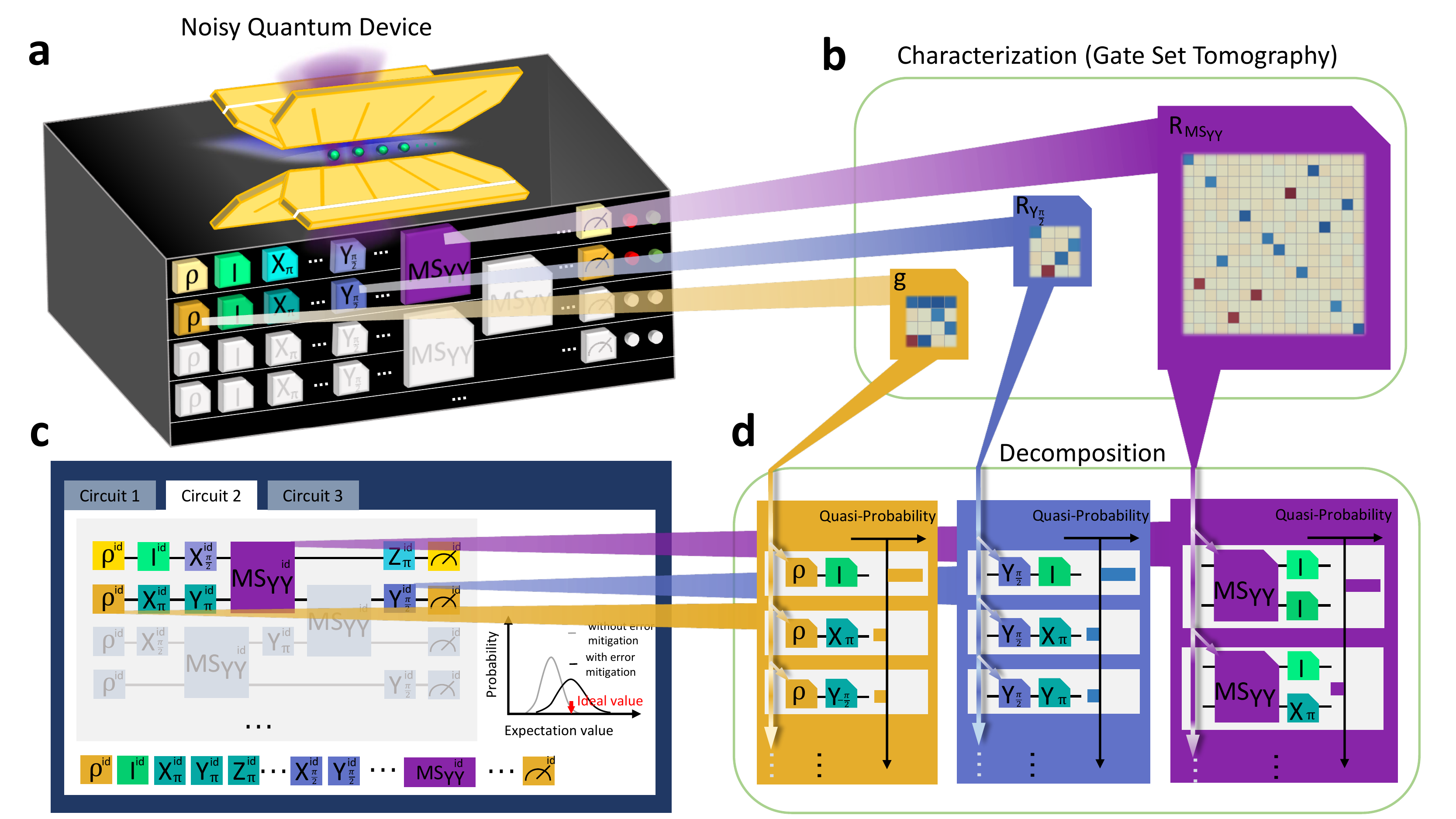}
		\caption{{\bf Paradigm of error-mitigated quantum computation.} {\bf a.} Quantum black box based on a trapped $\Yb$-ion system. Each button on the surface corresponds to an operation exerted on the quantum system encapsulated, where the buttons with $\rho$ and $M$ are for initial-state preparation and computational-basis measurement, whose results are indicated by the lights. The other buttons are for single-qubit and two-qubit quantum operations on certain qubits. The operations are implemented by global (blue) and individual (purple) laser beams illuminating the ion qubits. {\bf b.} Characterization of the quantum black box. The error-affected state preparation and measurement is characterized by the Gram matrix $g$, and the effect of each operation $G$, like $Y_{\frac{\pi}{2}}$ and ${\rm MS}_{YY}$, is described by a Pauli transfer matrix $R_G$ in the superoperator formalism, which is obtained by gate set tomography. {\bf c.} Construction of unbiased estimator of an expectation value specified by a quantum circuit, with building blocks including initial state preparation, different single-qubit and two-qubit gates, and final measurement. With error-mitigation, the distribution of the output expectation value is shifted towards the ideal value at a cost of enlarged variance. {\bf d.} Quasi-probability decomposition for the ideal initial state and exemplary single-qubit and two-qubit gates. Since the errors in state preparation and those in measurement are indistinguishable, we ascribe both of the errors to state preparation and decompose the ideal initial state with a set of experimental basis states, prepared by state initialization followed by a random fiducial gate. The noise part of an experimental quantum operations can be formally reversed by random experimental basis operations with quasi-probabilities obtained from decomposing the inverse of the noise operation with those of the experimental basis operations. Since the quasi-probabilities are real but can be negative, the random circuits can only be sampled in practice for the estimation of an expectation value, where the probabilities are proportional to the absolute values and the measurement outcomes are modified by the signs.}
	\end{figure*}
	
	Apart from physically improving the devices, the errors can be suppressed on the algorithmic level. For example, quantum error correction~\cite{shor1995scheme,steane1996error} provides a mean of fault-tolerant quantum computation. However, quantum error correcting codes require complex coding schemes, a large number of physical qubits and low error rates, which are still far from being affordable for near-term quantum technologies~\cite{o2017quantum,preskill2018quantum}. Consequently, it has not yet been demonstrated that quantum fault tolerance protocols can increase the fidelity of computation operations in any physical implementation. Alternatively, for the quantum algorithms estimating expectation values, the reliability of computation result can be improved by recently proposed error mitigation schemes~\cite{li2017efficient,temme2017error,endo2018practical,havlivcek2019supervised,kandala2019error} without challenging requirements for quantum error corrections. The probabilistic error cancellation method provides a comprehensive way to mitigate errors in expectation estimation tasks~\cite{temme2017error,endo2018practical,Song2018Quantum}. It begins with characterizing imperfect operations on the quantum device by tomography technique and then cancels errors by sampling random quantum circuits, according to a quasi-probability distribution derived from reconstructing ideal quantum operations with characterized imperfect ones.  Here we construct a trapped-ion system with full controllability and investigate the universal validity of the probabilistic error cancellation method in a general quantum computational context. We apply the method to every imperfect elementary quantum operation and benchmark the performance of error-mitigated quantum computation~\cite{knill2008randomized}. We observe dramatic improvements on effective fidelities of single- and two-qubit gates by an order of magnitude to those of physical gates.

	The paradigm of error-mitigated quantum computation is shown in Fig. 1. The noisy quantum device is treated as a multi-qubit black box in Fig. 1a, capable of preparing each qubit into an initial state $\rho_{0}$, performing a set of single-qubit and two-qubit gates and two-outcome measurement on each qubit, which is described by a positive operator-valued measure (POVM) ${\mathcal M}\equiv\set{E_0, {\mathbb I}-E_0}$ with ${\mathbb I}$ being the $2\times 2$ identity operator. These quantum operations are generally not accurate because of errors from operational imperfections and environmental fluctuations. As proposed in Ref.~\cite{endo2018practical}, we perform the gate set tomography~\cite{merkel2013self, blume2017demonstration, greenbaum2015introduction} and characterize state preparation and measurement (SPAM) and gates of noisy quantum devices by Gram matrices and Pauli transfer matrices (PTMs), respectively~\cite{greenbaum2015introduction} as shown in Fig. 1b. When we repeatedly execute a quantum circuit with such a noisy device aiming at obtaining the expectation values of observables of interest, the estimation will be deviated from the ideal case due to the imperfection of the quantum device, as shown in Fig. 1c. The correction of each noisy quantum operation can be decomposed to the combination of experimental basis operations (which we give later) with quasi-probabilities as shown in Fig. 1d. Since some of the quasi-probabilities can be negative, we cannot directly implement the decomposition. However, these basis operations can be randomly included in the circuit and resampled with the quasi-probabilities~\cite{temme2017error}. After running the random circuits with the corrections, the probability distribution of the output expectation value is shifted towards the ideal value at a cost of enlarged variance due to the presence of negative values in the quasi-probabilities~\cite{endo2018practical}, as shown in Fig. 1c. The variance can be reduced by increasing the repetition number, which is the number of generated random circuits.
	
		\begin{figure*}[ht]
		\includegraphics[width = \textwidth]{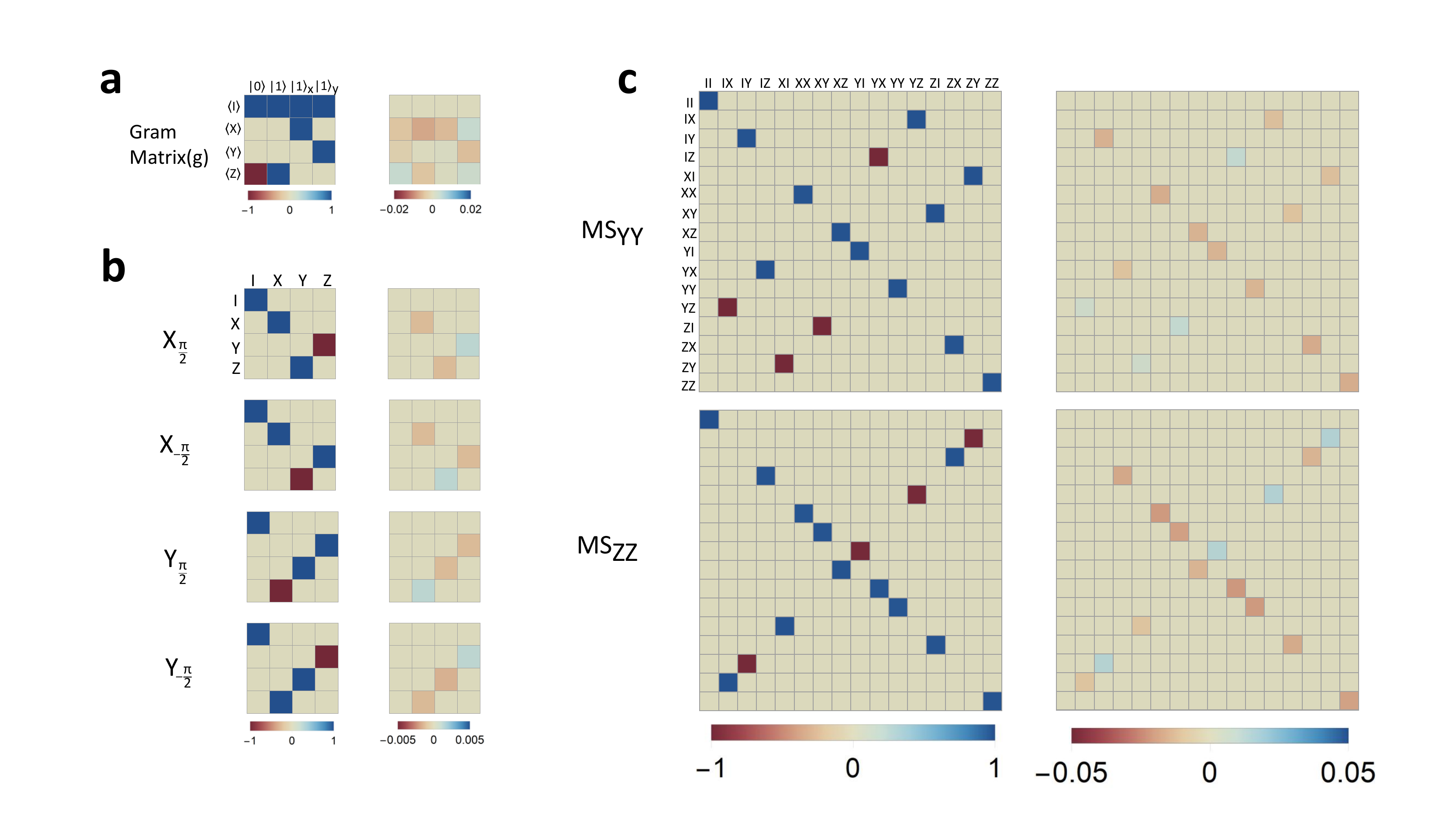}	
		\caption{{\bf Characterization of noisy quantum devices obtained by gate set tomography.} {\bf a.} Gram matrix of the single-qubit case. The Gram matrix characterize the SPAM error, which is obtained by preparing states in ${\mathcal S}_1$ and measuring expectation values of operators in ${\mathcal P}_1$. {\bf b.} Pauli transfer matrices for the experimental realizations of the $X_{\pm\frac{\pi}{2}}$ and $Y_{\pm\frac{\pi}{2}}$ gates for the single-qubit gates. {\bf c.} Pauli transfer matrices of the experimental realizations of the ${\rm MS}_{YY}$ and  ${\rm MS}_{ZZ}$ gates in the two-qubit case. In each subfigure, the left column shows the experimentally-obtained matrices and the right column shows the difference between the experimental and the ideal matrices, i.e. $R_G-R_G^{\rm id}$ with $G$ being one of the quantum operations being characterized.}
	\end{figure*}
	
	In our experimental realization, the quantum hardware encapsulated in the black box is a trapped-ion system, where $^{171}{\rm Yb}^+$ ions are trapped into a linear crystal and individually manipulated by global and individual laser beams, as shown in Fig. 1a. To encode quantum information, a pair of clock states in the ground-state manifold $^2S_{1/2}$, i.e. $\ket{F=0,m_F=0}$ and $\ket{F=0,m_F=1}$, are denoted as the computational basis $\left\{\ket{0},\ket{1}\right\}$ of a qubit. At the beginning of executing a quantum circuit, each ion qubit is initialized to $\ket{0}$ by optical pumping. We implement single-qubit operations by Raman laser beams with beatnote frequency about the hyperfine splitting $\omega_0=2\pi\times12.642821~{\rm GHz}$. And the two-qubit operation, i.e. the M\o lmer-S\o rensen $YY$-gate (${\rm MS}_{YY}$) is realized by driving transverse motional modes\cite{sorensen1999quantum, kim2009entanglement}, with frequencies in the x-direction \{$\nu_1$, $\nu_2$\}=\{1.954, 2.048\} MHz. We apply amplitude-shaped ~\cite{schindler2008} bichromatic Raman beams with beatnote frequencies  $\omega_0\pm\mu$, where $\mu$ is set to be the middle frequency of the two motional modes, and achieve the ${\rm MS}_{YY}$ gate for 25 $\mu$s. We also realize the M\o lmer-S\o rensen $ZZ$-gate (${\rm MS}_{ZZ}$) by adding single-qubit rotations before and after the ${\rm MS}_{YY}$ gate \cite{tan2015multi}. At the end of the execution, internal states of qubits are measured by state-dependent fluorescence detection~\cite{olmschenk2007manipulation}. Note that to collect fluorescence photons, we use a photomultiplier tube (PMT) in the single-qubit case and an electron-multiplying charge-coupled device (EMCCD) in the two-qubit case.
	
		\begin{figure*}[ht]
		\includegraphics[width = \textwidth]{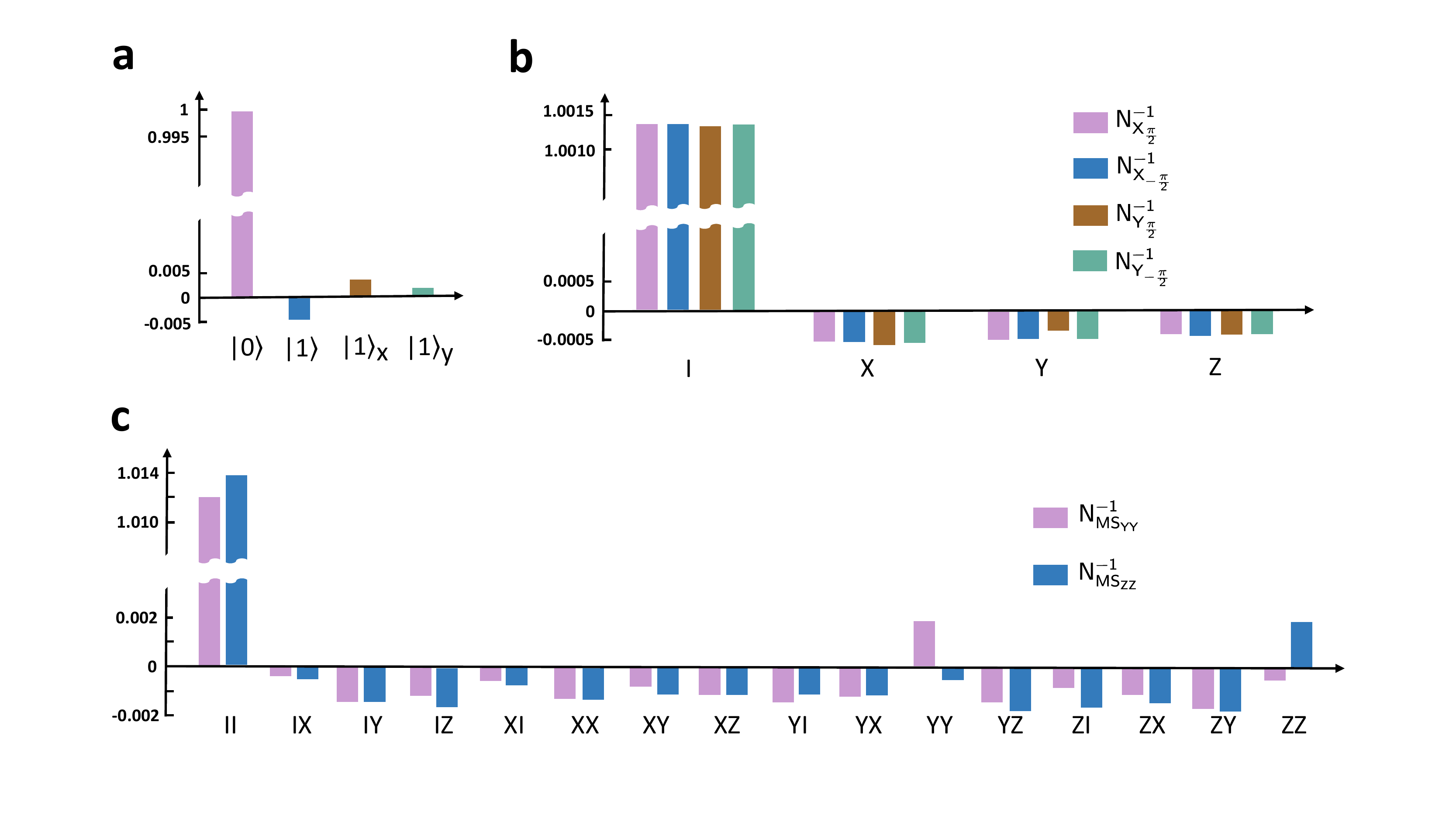}	
		\caption{{\bf Quasi-probability decomposition.} {\bf a.} Quasi-probabilities in the decomposition of the ideal single-qubit initial state with experimental initial states in ${\mathcal S}_1$. {\bf b.} Quasi-probabilities in the decomposition of the inverse noise operations of the four experimental single-qubit gates $\set{X_{\pm\frac{\pi}{2}}, Y_{\pm\frac{\pi}{2}}}$. {\bf c.} The same as {\bf b.} for the experimental two-qubit gates $\set{{\rm MS}_{YY}, {\rm MS}_{ZZ}}$.}
	\end{figure*}
	
	We introduce the PTM representation for the mathematical description of an $n$-qubit noisy quantum device, where density operators $\rho$ and physical observable $E$ are represented by $2^n$-entry column vectors $\colvec{\rho}$ and row vectors $\rowvec{E}$, and quantum gates $G$ are represented by $2^{2n}\times 2^{2n}$ PTMs $R_G$. Here, the expectation value of the observable $\hat{E}$  after operating $G_s$ on the initial state $\hat{\rho}$ is represented by $\hsmatele{E}{ R_G}{\rho}$. PTMs can be determined by gate set tomography, which  requires informationally complete data obtained from experiments with initial states from a basis set ${\mathcal S}_n \equiv \set{\ket{0}, \ket{1}, \ket{1}_X, \ket{1}_Y}^{\otimes n}$ and measurement of the observables from the $n$-qubit Pauli basis ${\mathcal P}_n=\set{{\mathbb I}, X, Y, Z}^{\otimes n}$. Compared to quantum process tomography, gate set tomography is featured by appropriately taking consideration of  SPAM errors, which is of great importance in quantum computations with high accuracy. In gate set tomography, the states in $\mathcal{S}_n$ and the  measurement of observables in $\mathcal{P}_n$ are realized by using a set of fiducial gates $\mathcal{F}_n \equiv \set{{\mathbb I}, X_\pi, Y_{-\frac{\pi}{2}}, X_{\frac{\pi}{2}}}^{\otimes n}$ consisting of the identity operation and the $X$ or $Y$ axis rotations on each qubit, which are to be characterized together with the rest of the quantum operations. The single-qubit SPAM errors are reflected in the Gram matrix~\cite{greenbaum2015introduction}, as shown in Fig. 2a, which is obtained by preparing the qubit in one of the states ${\mathcal S}_1$, $\colvec{\rho_i}=R_{F_i} \colvec{\rho_0}$, and measuring the expectation values of the operators in the single-qubit Pauli basis ${\mathcal P}_1$, $\rowvec{E_i}=\rowvec{E_0}R_{F_i}$, where $\rho_0$ and $E_0$ are ideally associated with $\ket{0}\bra{0}$ and $Z$, respectively.
	
	For single-qubit randomized benchmarking \cite{knill2008randomized}, we design pulse sequences for implementing major-axis $\pi$ pulses $\set{X_{\pm\pi}, Y_{\pm\pi}, Z_{\pm\pi}}$ and $\frac{\pi}{2}$ pulses $\set{X_{\pm\frac{\pi}{2}}, Y_{\pm\frac{\pi}{2}}}$. Thus the gate set for the single-qubit case is ${\mathcal G}_1 = \set{{\mathbb I}, X_{\pm\pi}, Y_{\pm\pi}, Z_{\pm\pi}, X_{\pm\frac{\pi}{2}}, Y_{\pm\frac{\pi}{2}}}$, where ${\mathbb I}$ is the identity operation. The gate set for implementing two-qubit random circuits are ${\mathcal G}_2={\mathcal G}_1^{\otimes 2}\cup\set{{\rm MS}_{YY}, {\rm MS}_{ZZ}}$. After obtaining informationally complete data for each quantum operation, PTM can be obtained by the linear inversion GST in principle~\cite{endo2018practical}. However, the accuracy of this approach is highly affected by the sampling error in practice, which results in unphysical PTMs. Therefore, we use maximum likelihood estimation for the reconstruction of PTM. We also assume Pauli errors are dominant in our device, where each of the noisy quantum gate $G_s\in{\mathcal G}_n$ is modeled with the ideal gate $G^{\rm id}_s$ followed by a Pauli error channel $\Lambda_s$, i.e. $G_s=\Lambda_s\circ G_s^{\rm id}$. The Pauli error channel, $\Lambda_s(\rho) =\sum_j p_{s,j} P_j \rho P_j $ where $P_j$ are Pauli operators, is characterized by a set of Pauli error rates $ p_{s,j}$, which are determined by maximizing the likelihood function defined as follows,
	\begin{eqnarray}
	{\mathcal L}= \prod_{i,j,k}\exp\left[-(m_{ijk}-\bar m_{ijk})^2/\Delta_{ijk}^2\right],
	\end{eqnarray}
	where $m_{ijk}$ and $\bar m_{ijk}$ are ansatz prediction and experimentally observed frequency of the probability $\hsmatele{E_i}{R_{G_j}}{\rho_k}$, respectively. Here, the POVM element is $\rowvec{E_i}\equiv\rowvec{E_0}R_{F_i}\equiv\rowvec{0}R_{F_i}$, and $\Delta_{ijk}$ is the standard deviation of the experimental data for $\bar m_{ijk}$. 
	
	\begin{figure*}[ht]
		\includegraphics[width = \textwidth]{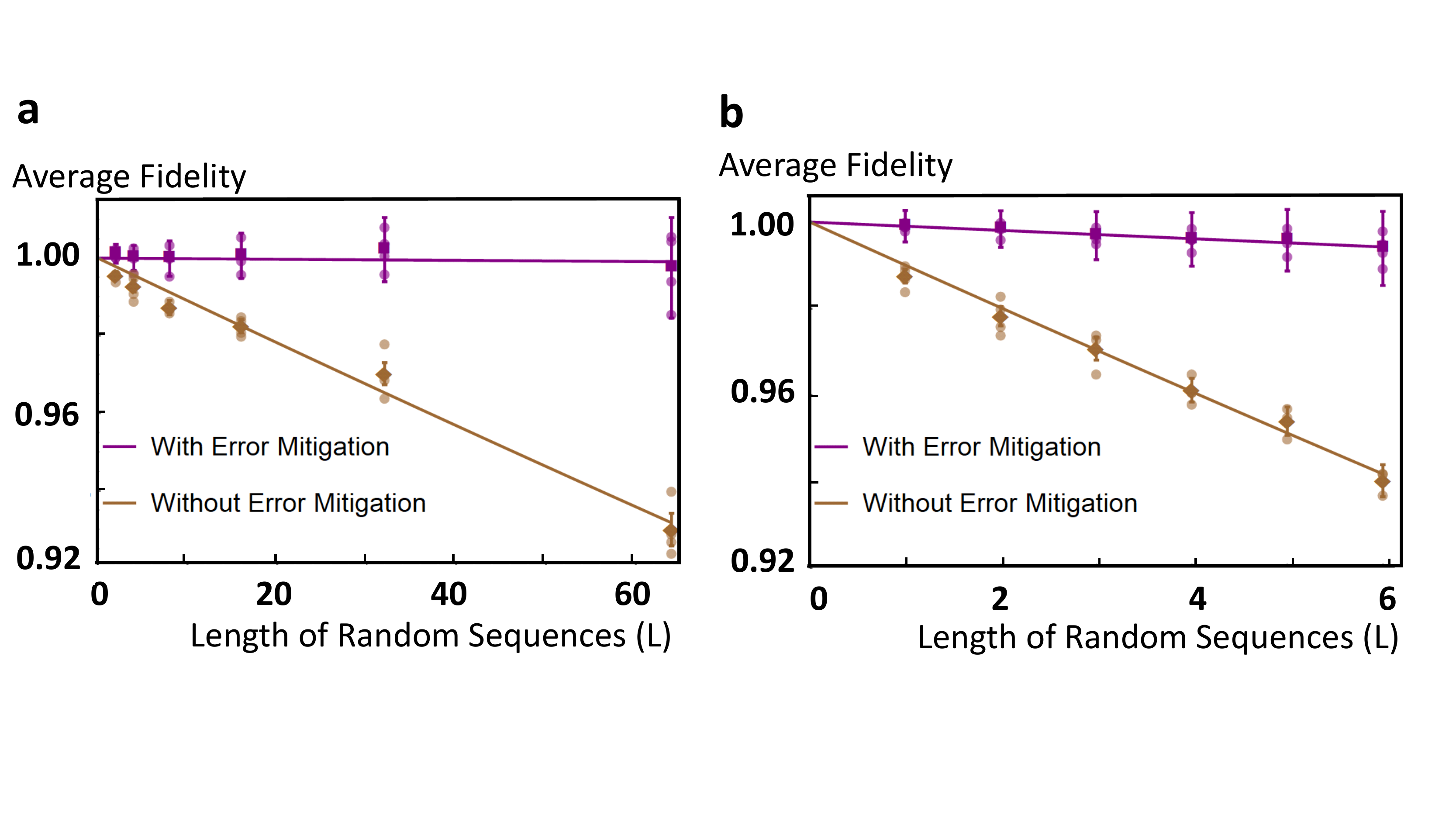}	
		\caption{{\bf Experimental data for error-mitigated quantum computation.} {\bf a.} The single-qubit randomized benchmarking. The data points represent the average fidelity for random sequences of length $L$, and the curves are obtained from numerical exponential fitting. The decay rate per gate indicated by the experimental randomized benchmarking curve (yellow) is $(1.10\pm 0.12)\times10^{-3}$, while after error-mitigation, the effective decay rate is suppressed to $(1.44\pm 5.28)\times10^{-5}$. {\bf b.} The two-qubit random-circuit computation. Decay rates indicated by the average fidelity curves without and with error mitigation are $(0.99\pm 0.06)\times10^{-2}$ and $(0.96\pm 0.10)\times10^{-3}$, respectively.}
	\end{figure*}
	
	The reconstructed PTMs of $X_{\pm\frac{\pi}{2}}$ and $Y_{\pm\frac{\pi}{2}}$ for the single-qubit case and those of ${\rm MS}_{YY}$ and ${\rm MS}_{ZZ}$ gates for the two-qubit case are shown in Fig. 2b and Fig. 2c, respectively (more data for the single-qubit case in Supplementary Fig. S1a) . We note that, for the gate set tomography of two qubits, we apply a two-step parameter estimation, since the infidelities for the single-qubit gates are about an order lower than those of the two-qubit gates. We first determine the Pauli error rates for all the single-qubit gates in ${\mathcal G}_1^{\otimes 2}$ as described above, and then characterize the two-qubit gate ${\rm MS}_{YY}$ based on the knowledge of the characterized single-qubit gates (see Supplementary Materials). The  ${\rm MS}_{ZZ}$ gate is derived from those results. Using these reconstructed PTMs, we simulate the single-qubit randomized benchmarking and two-qubit random circuits. The comparisons  between the reconstructed and experimental data clearly validate the Pauli error assumption within both error-bars (see Supplementary Fig. S2).  
	
	The initial state, quantum gates and measurement are deviated from the ideal ones, as experimentally characterized by Gram matrix and PTMs. Mathematically, We can reconstruct the ideal ones by weighted combination of experimental operations~\cite{temme2017error}. Since we cannot distinguish errors in state preparation from those in measurement, we ascribe all of the SPAM errors to state preparation and decompose the initial state $\colvec{\rho_0^{\rm id}}=\sum_i q_{0,i}\colvec{\rho_i}$.  The quasi-probabilities $q_{0,i}$ for the decomposition of the ideal single-qubit initial state is shown in Fig. 3a. Note that for the two-qubit case, the SPAM errors are much more serious because of the EMCCD, and we calibrate the results to remove the SPAM errors as proposed in Ref.~\cite{shen2012correcting}.
	
	An ideal quantum gate $G_s^{\rm id}$ can be written as the experimental one followed by the inverse of noise operation,  i.e. $R_{G_s}^{\rm id}=N_s^{-1} R_{G_s}$, where the noise operation $N_s$ introduces errors in the experimental gate $R_{G_s}=N_s R_{G_s}^{\rm id}$. The inverse of the noise operation $N_s^{-1}$ is then decomposed by the experimental operations associated with the $n$-qubit Pauli group, $N_s^{-1}=\sum_jq_{s,j}R_{P_j}$ with Pauli error assumption, where the quasi-probabilities $q_{s,j}$ are determined by a set of linear equations. We show decompositions of the inverse error operations for single-qubit gates $\set{X_{\pm\frac{\pi}{2}},Y_{\pm\frac{\pi}{2}}}$ in Fig. 3b (more data in Supplementary Fig. S1b) and for two-qubit gates $\set{{\rm MS}_{YY},{\rm MS}_{ZZ}}$ in Fig. 3c. 
	
	To benchmark the performance, we implement the error-mitigated single-qubit and two-qubit random circuits, both of which consist of implementing a number of randomized computational sequences on fully polarized initial states, $\ket{0}$ and $\ket{00}$, and measuring $Z$ on each qubit. For the single-qubit case, a random sequence of length $L$ for randomized benchmarking contains $L$ computational gates and $L+1$ interleaving identity or Pauli operations, uniformly drawn from the set $\set{X_{\pm\frac{\pi}{2}}, Y_{\pm\frac{\pi}{2}}}$ and $\set{{\mathbb I}, X_{\pm \pi}, Y_{\pm\pi}, Z_{\pm\pi}}$, respectively. We post-select the uniformly-generated random sequences to minimize the projection error, and choose those with ideal final states being an eigenstate of the Pauli $Z$ operator. In order to obtain unbiased estimator of the expectation value, both the experimentally prepared initial state and the $2L+1$ experimental operations need to be decomposed and resampled, where the initial state is replaced probabilistically by one of the states in ${\mathcal S}_1$, and each experimental gate is followed by a random Pauli or identity operation drawn from ${\mathcal P}_1$. Thus, for a random sequence of length $L$, there are $4^{2L+2}$ possible experimental settings. Since the number of settings grows exponentially with the length of the random sequence, we use Monte-Carlo sampling to generate experimental settings, which are specified by an index $i$ for the initial state $\colvec{\rho_i}$ and two $(2L+1)$-entry index vectors ${\mathbf a}$ and ${\mathbf b}$ specifying the target sequence and the choices of the error-compensating operations. The probability of an experimental setting $\hsmatele{E_0}{\prod_{l=1}^{2L+1}R_{P_{b_l}}R_{G_{a_l}}}{\rho_i}$, where $G_{a_l}\in{\mathcal G}_1$ and $P_{b_l}\in{\mathcal P}_1$, is $C^{-1}\left|q_{0,i}\left(\prod_{l=1}^{2L+1}q_{a_l,b_l}\right)\right|$. Here, the random variable ${\mathbf a}$ is introduced to benchmark the performance, the random variable ${\mathbf b}$ is used to cancel errors, and the normalization constant $C=\sum_{i,\ldots,\left(a_l,b_l\right),\ldots}\left|q_{0,i}\left(\prod_{l=1}^{2L+1}q_{a_l,b_l}\right)\right|\geq 1$ characterizes the cost to mitigate the errors. Note that the signs of the coefficients, i.e., ${\rm sgn}\left[q_{0,i}\left(\prod_{l=1}^{2L+1}q_{a_l,b_l}\right)\right]$, are integrated into the measurement results of the random experiments. In Fig. 4a, we represent the error-mitigated single-qubit randomized benchmarking with length $L$ up to $64$, and show that the single-qubit gate error rate is effectively suppressed from $(1.10\pm 0.12)\times10^{-3}$ to $(1.44\pm 5.28)\times10^{-5}$. 
		
	For the two-qubit case, we first uniformly generate a number of random sequences, and after post-selection, end up with $4$ random sequences for each sequence length $L$, whose ideal final states are an eigenstate of $Z^{\otimes 2}$. Here, a random sequence of length $L$ contains $L$ two-qubit gates uniformly drawn from the set $\set{{\rm MS}_{YY}, {\rm MS}_{ZZ}}$, which is then randomized by interleaving the two-qubit gates with random single-qubit gates \cite{gaebler2012randomized}. Similar to the single-qubit case described above, we apply error mitigation to each of the two-qubit random sequences with length $L$ up to 6, and represent the error-mitigated results in Fig. 4b, where the two-qubit gate error rate is effectively suppressed from $(0.99\pm 0.06)\times10^{-2}$ to $(0.96\pm 0.10)\times10^{-3}$.  	
	Our work shows that the error mitigation technique, i.e. probabilistic error cancellation~\cite{temme2017error, endo2018practical, Song2018Quantum}, surely have the capacity of surpassing the break-even point, where the effective gates are superior to their physical building blocks, at an affordable cost with respect to near-future quantum techniques. The effective infidelity after error mitigation comes from the Pauli error assumption, time-dependent systematic drifting~\cite{mavadia2018experimental} for both single-qubit and two-qubit cases and cross-talk error of single-qubit addressing operations for the two-qubit case. Thus further improvement requires both calibrating and stabilizing the quantum device. With technologies to tackle the cross-talk error, the probabilistic error cancellation method of quantum error mitigation can be straightforwardly applied to systems with more qubits for realizing high-fidelity quantum computation.

	\section*{Acknowledgments}
	This work was supported by the National Key Research and Development Program of China under Grants No. 2016YFA0301900 and No. 2016YFA0301901 and the National Natural Science Foundation of China Grants No. 11574002, No. 11504197, and No. 11875050. Y. L. also acknowledges the support by NSAF Grant No. U1730449.
	
	\beginsupplement
	\section*{Supplementary materials}

	\subsection*{Characterization and decomposition for single-qubit operation with Pauli error assumption}
	
	We use gate set tomography to characterize the single-qubit operations. In the superoperator formalism, each experimental single-qubit operation $R_{G_s}$ can be describe as an ideal 4 by 4 PTM followed by a PTM of noise operation $N_s$. With Pauli error assumption, each $N_s$ can be written as $N_s=p_{s,0}R_{\mathbb{I}}+p_{s,1}R_X+p_{s,2}R_Y+p_{s,3}R_Z$, where $p_{s,j}$ are the Pauli error rates and $\sum_j p_{s,j}=1$ for trace preserving condition. Since there are 11 gate in ${\mathcal G}_1$, ${\mathcal F}_1 \subset {\mathcal G}_1$ and the experimental initial state $\rho_0$ can be characterized by 3 parameters, we need to obtain the values for $11 \times 3 +3=36$ parameters. We run $3\times 11 \times 4$ different experimental settings specified by  $\hsmatele{E_0}{R_{F_k}R_{G_j}R_{F_i}}{\rho_0}$ with repetitions of 10000 per setting to collect experimental data $\bar m_{ijk}$, where $i=1,\ldots, 4$ for state preparation, $j=1,\ldots, 11$, and $k=1,2,3$ for different measurement settings. The ansatz prediction $m_{ijk}=\hsmatele{E_0}{N_{F_k}R^{id}_{F_k}N_{G_j}R^{id}_{G_j}N_{F_i}R^{id}_{F_i}}{\rho_0}$ contain Pauli error rates as variational parameters, which we numerically optimize to maximize the likelihood function in Eq.(1). The obtained PTMs are shown in Fig. 2b and Fig. S1a.
	
	Once we get experimental PTMs  for single-qubit operations, we can derive the inverse of PTM of the noise operation as $N^{-1}_s=R^{id}_{G_s}R^{-1}_{G_s}$
	, which can be decomposed by the combination of PTMs of experimental Pauli operations with 
	$N^{-1}_s=q_{s,0}R_{\mathbb I}+q_{s,1}R_{X}+q_{s,2}R_{Y}+q_{s,3}R_{Z}$. 
	Then, the ideal operation can be decomposed by experimental operations as 
	$$R^{id}_{G_s}=q_{s,0}R_{\mathbb I}R_{G_s}+q_{s,1}R_XR_{G_s}+q_{s,2}R_YR_{G_s}+q_{s,3}R_ZR_{G_s}.$$
	
		\begin{figure*}[ht]
		\includegraphics[width = \textwidth]{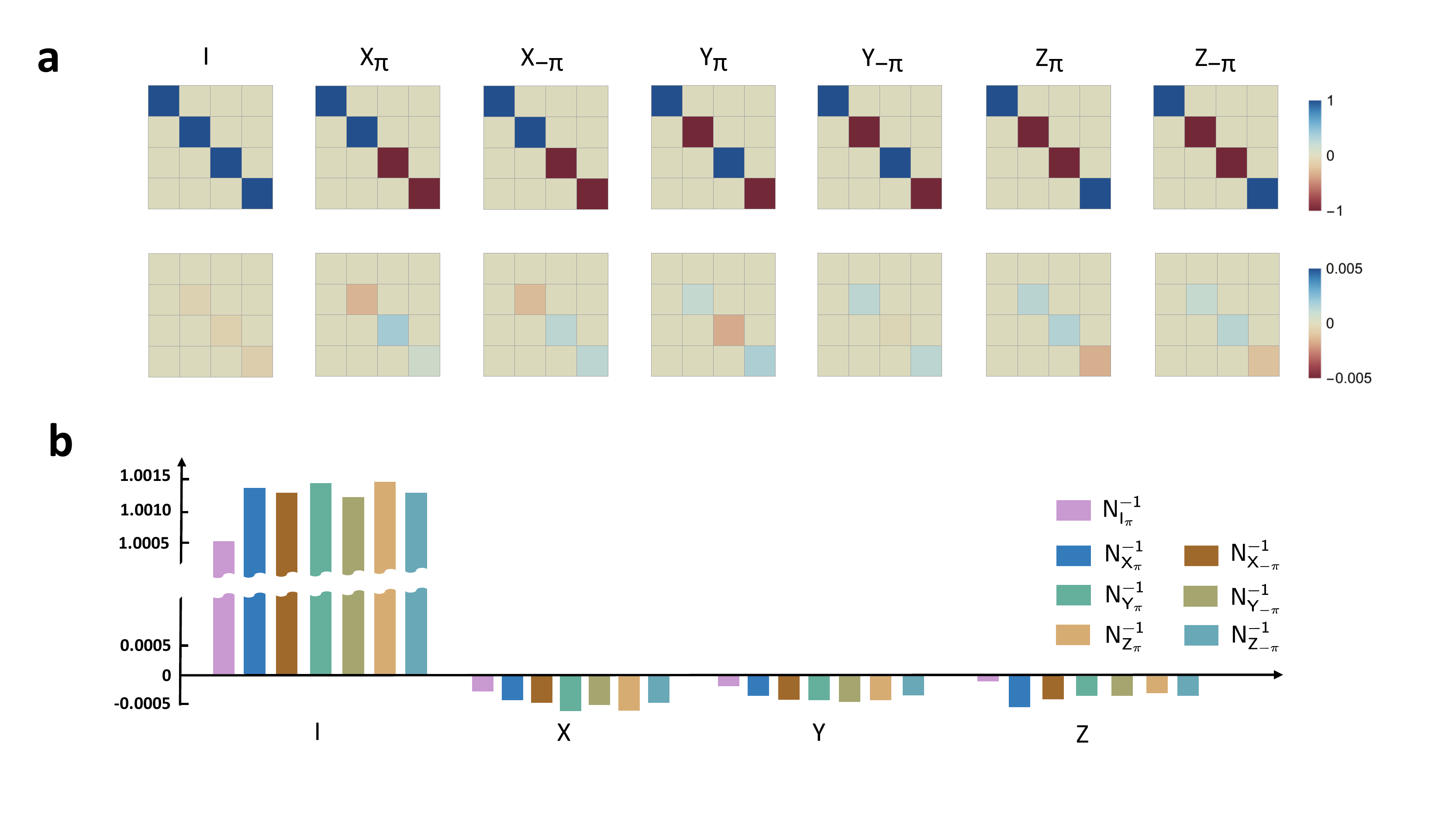}
		\caption{{\bf Characterization and decomposition of the experimental identity and $\pi$ pulses.} {\bf a.} Single-qubit randomized benchmarking needs not only the computational gate set, but also the identity and $\pi$ pulses~\cite{knill2008randomized}. In order to implement error-mitigation for single-qubit randomized benchmarking, these gates should be characterized and the errors should also be decomposed. The upper row shows the experimentally-obtained PTMs and the lower row shows the difference between the experimental and the ideal matrices. {\bf b.} Quasi-probabilities in the decomposition of the inverse noise operations of these single-qubit gates. }
	\end{figure*}
	
		\begin{figure*}[ht]
		\includegraphics[width = \textwidth]{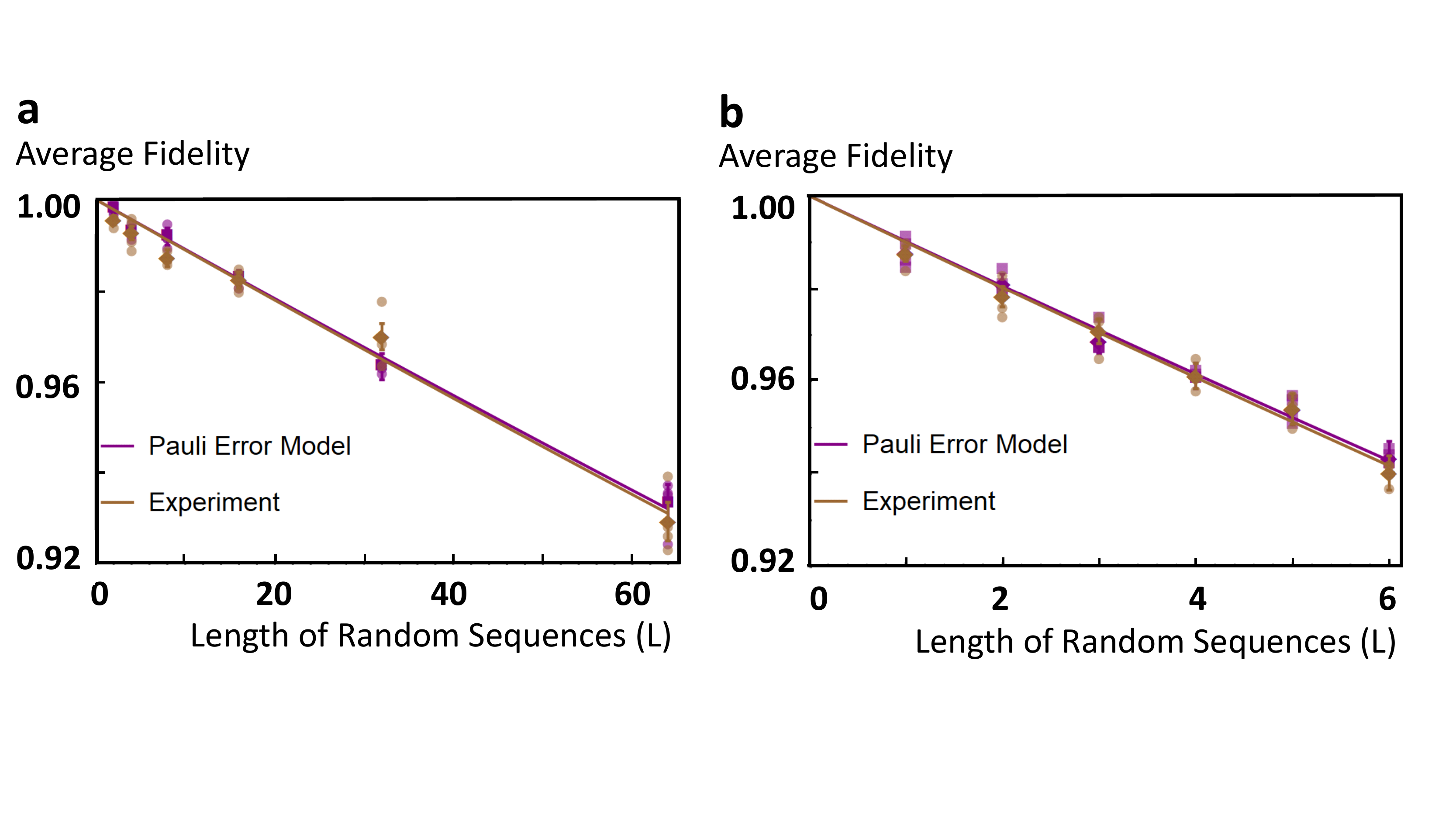}
		\caption{{\bf Verification of the Pauli-error assumption.} {\bf a}. The average fidelity of the numerical (purple dots) and experimental (yellow dots) single-qubit random sequences as functions of the sequece length $L$. The numerical data are obtained by simulating the quantum dynamics with the experimental PTMs with the Pauli-error assumption. The curves are obtained by fitting an exponential decaying model to the data. The numerical and experimental error rates, being $(1.09\pm 0.06)\times10^{-3}$ and $(1.10\pm 0.12)\times10^{-3}$ respectively, are consistent within fitting errors. {\bf b}. The same as {\bf a}. for random two-qubit sequences. The numerical and experimental error rates are $(0.97\pm 0.05)\times10^{-2}$ and $(0.99\pm 0.06)\times10^{-2}$. Thus the comparison in {\bf a}. and {\bf b}. validate the Pauli-error assumption in our system.}
	\end{figure*}
	
	\subsection*{Characterization of the two-qubit gate set}
	
	The two-qubit gate set, i.e. ${\mathcal G}_2$, includes single-qubit operations in ${\mathcal G}_1^{\otimes2}$ and two-qubit operations $\set{{\rm MS}_{YY}, {\rm MS}_{ZZ}}$. Since infidelities for the single-qubit gates are about an order lower than those of the two-qubit gates, it is reasonable to divide the maximum likelihood estimation into two steps. 
	
	First, we treat each qubit in the two-qubit system as a single-qubit system and characterize the single-qubit gate set ${\mathcal G}_1$ by gate set tomography, obtaining single-qubit PTMs. The two-qubit PTMs of the single-qubit operations in ${\mathcal G}_1^{\otimes2}$ is then obtained by direct product of the single-qubit PTMs on both qubits. Since the fiducial set ${\mathcal F}_2\in{\mathcal G}_1^{\otimes 2}$, the PTMs of the fiducial operations are determined at this step.
	
	Second, we characterize the native two-qubit ${\rm MS}_{YY}$ gate. Under the Pauli-error assumption, the PTM of the experimental ${\rm MS}_{YY}$ gate is decomposed as $R_{{\rm MS}_{YY}} = N_{{\rm MS}_{YY}}R^{\rm id}_{{\rm MS}_{YY}}$, where $N_{{\rm MS}_{YY}}$ is the PTM of the Pauli-error channel containing $16$ two-qubit Pauli components. After considering the trace-preserving constraint, $N_{{\rm MS}_{YY}}$ has 15 parameters, which are determined by linear equations connecting the ansatz predition $\hsmatele{E_0^{\otimes 2}}{R_{F_k}N_{{\rm MS}_{YY}}R^{\rm id}_{{\rm MS}_{YY}}R_{F_i}}{\rho_{0}^{(1)}{\otimes }\rho_0^{(2)}}$ and corresponding experimental results. In order to minimize the projection error, we choose $15$ linearly independent equations out of $16\times9$ different settings, with most of the measured probabilities close to $0$ or $1$. Fig. S3 shows the corresponding circuits for the experimental settings.
	
	Since the ${\rm MS}_{ZZ}$ is implemented by a ${\rm MS}_{YY}$ gate sandwiched by proper single-qubit gates, the PTM of the experimental ${\rm MS}_{ZZ}$ gate is obtained by multiplying the PTMs of the corresponding experimental operations, i.e. $R_{{\rm MS}_{ZZ}}=R_{X_{-\frac{\pi}{2}}\otimes {X_{-\frac{\pi}{2}}}}R_{{\rm MS}_{YY}}R_{X_{\frac{\pi}{2}}\otimes {X_{\frac{\pi}{2}}}}$.	
	
	
		\begin{figure*}[ht]
		\includegraphics[width = \textwidth]{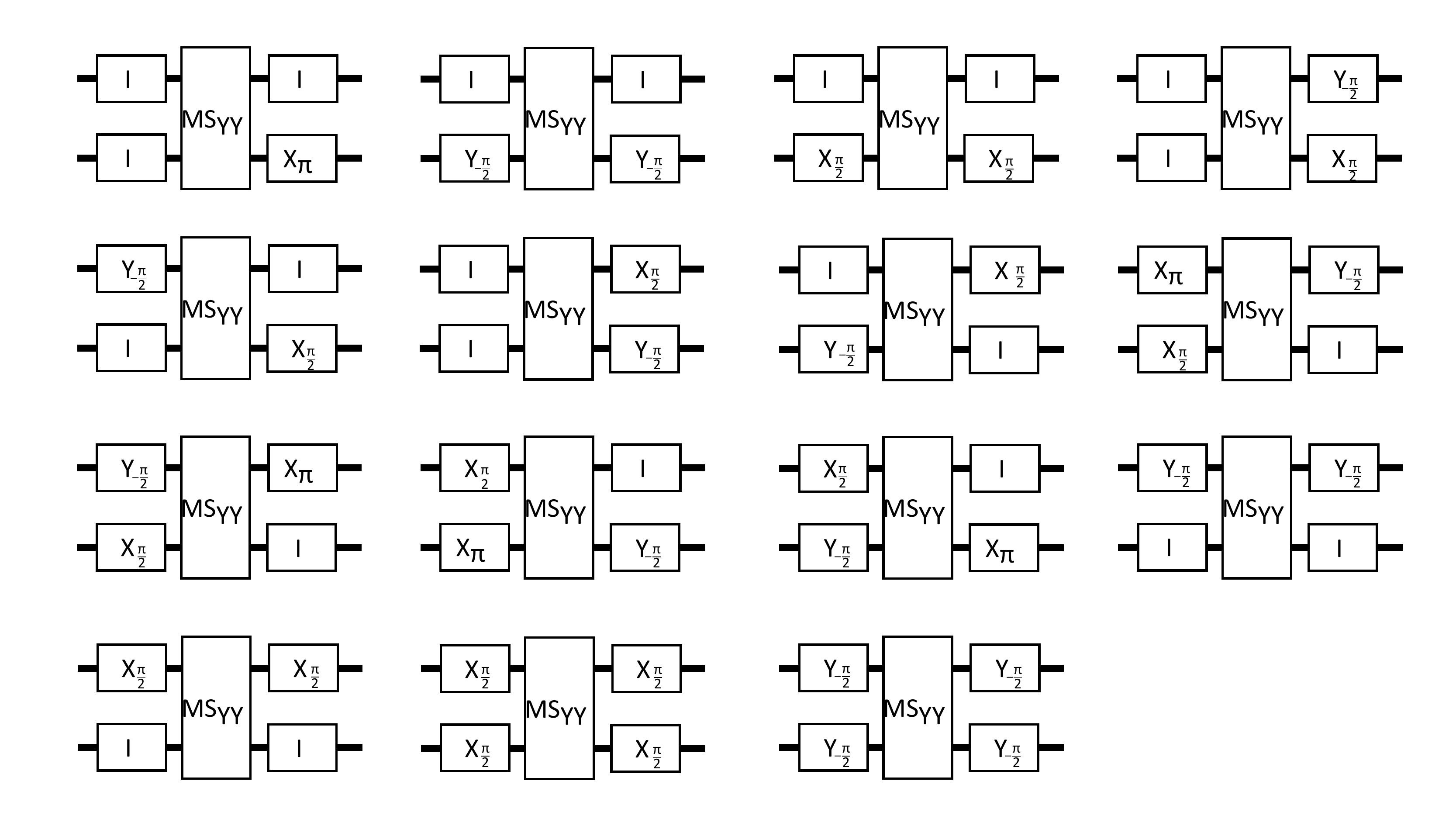}
		\caption{{\bf Experimental circuits for the characterization of the ${\rm MS}_{YY}$ gate.} The two-qubit system is first prepared in the initial $\ket{00}$ state by optical pumping. After implementing one of the quantum circuit, a projective measurement of $Z^{\otimes2}$ is carried out. The above sequence is repeated $3000$ times for each circuit to estimate the probability of the dark $\ket{00}$ state, which, together with the corresponding ansatz prediction, determines one of the Pauli-error rate for the experimental ${\rm MS}_{YY}$ gate.}
	\end{figure*} 
	
	\subsection*{Analysis on residual errors in the error-mitigated quantum computation}
	
		\begin{figure*}[ht]
		\includegraphics[width = \textwidth]{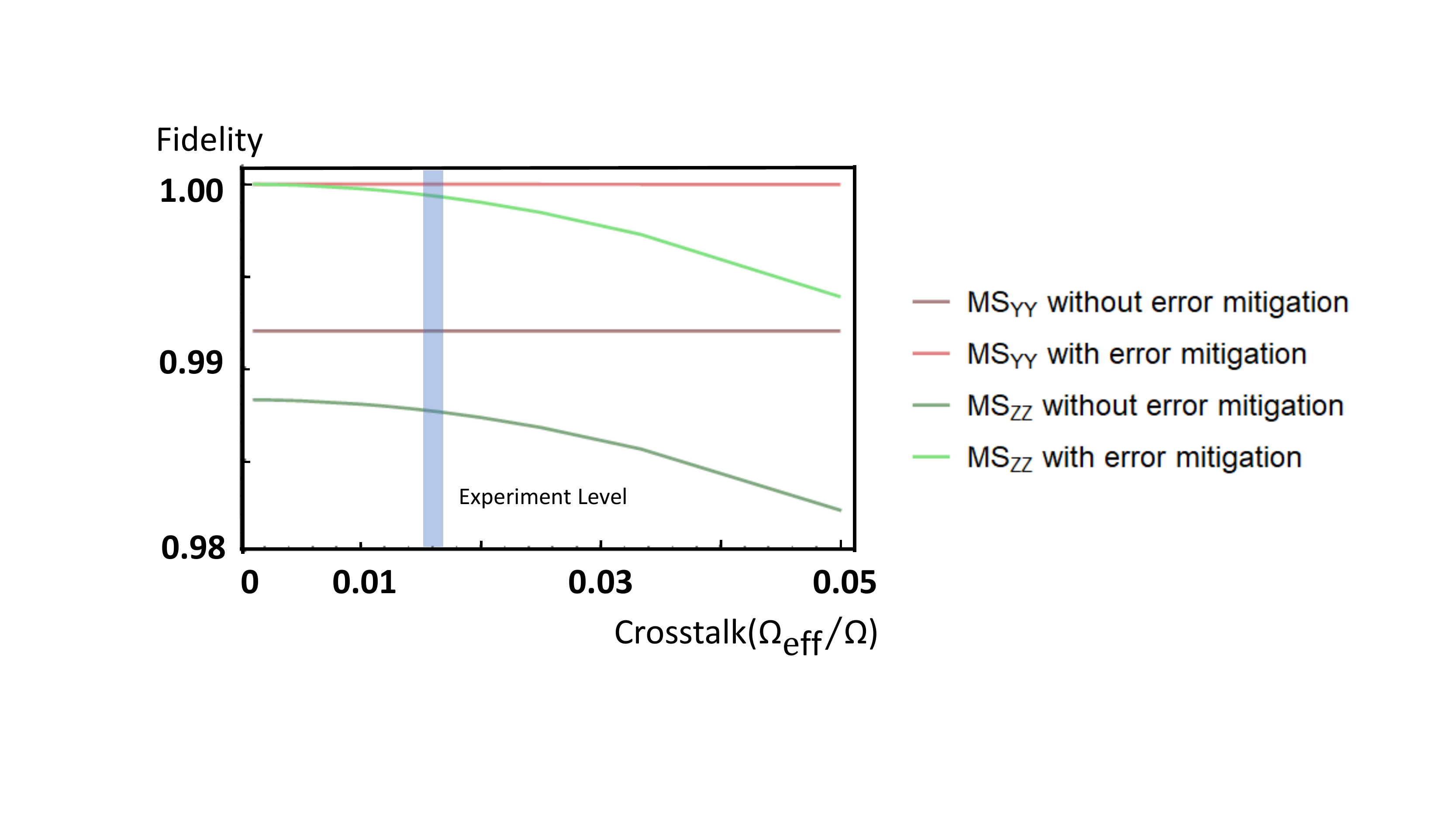}
		\caption{{\bf Analysis of the qubit crosstalk effect.} We numerically simulate the final-state fidelity of the original and error-mitigated ${\rm MS}_{YY}$ and ${\rm MS}_{ZZ}$ gates as functions of the qubit crosstalk strength, which is modeled by the ratio $\Omega_{\rm eff}/\Omega$, with $\Omega$ and $\Omega_{\rm eff}$ being the Rabi frequencies experienced by the target and the neighboring ions when a single-qubit gate is being implemented. The experimental level of the qubit crosstalk strength is shaded with blue, which given an estimation of $0.68\times10^{-3}$ for the residual error rate induced by the qubit crosstalk effect.}
	\end{figure*}
	
	Theoretically, the error mitigation technique, combining probabilistic error cancellation and gate set tomography, is capable of completely rectifying the effect of errors in the estimation of expectation values. However, in our experiment, the effective error rates after error mitigation are $(1.44\pm 5.28)\times 10^{-5}$ and $(0.96\pm 0.10)\times 10^{-3}$ in the single-qubit and two-qubit cases, respectively. Generally speaking, the reasons for the residual errors include the Pauli-error assumption, time-correlated systematic drift, and crosstalk errors between qubits.
	
	In the single-qubit case, the residual errors mainly come from the introduction of the Pauli-error model. To quantify the non-Pauli error rate, we simulate the dynamics of the same random sequences as those used in the experiment with the characterized experimental PTMs, which are obtained under the Pauli-error assumption. The experimental and simulated data of average fidelity are shown in Fig. S2a, which are then numerically fitted to extract the error rates. The difference between the simulated and experimental error rates for single-qubit gates is $1.41\times 10^{-5}$, which are of the same order of the residual error rate in the single-qubit case. Meanwhile, the data shows that the time-correlated systematic drift has negligible effect and cannot be faithfully quantified within experimental and fitting errors. 
	
	In our experiment, we implement two different two-qubit gates, i.e. ${\rm MS}_{YY}$ and ${\rm MS}_{ZZ}$ gates. To quantify the residual errors from the Pauli-error assumption, we compare the dynamics of the simulated and experimental random two-qubit sequence, where the simulation is based on the characterized PTMs with the Pauli-error assumption. The difference between the simulated and experimental error rates gives the estimation of the non-Pauli residual error rate, which is about $0.20\times 10^{-3}$. As to the crosstalk errors, the situations for ${\rm MS}_{YY}$ and ${\rm MS}_{ZZ}$ gates are quite different because of different implementation schemes. Specifically, a ${\rm MS}_{ZZ}$ gate is implemented by a ${\rm MS}_{YY}$ gate sandwiched by proper single-qubit gates, which introduce qubit-crosstalk errors. We model the crosstalk effect by introducing an effective Rabi frequency $\Omega_{\rm eff}$ on the neighboring ion induced by leakage laser intensities when a single-qubit gate is being implemented by lasers focused on one of the ions. The ratio $\Omega_{\rm eff}/\Omega$, with $\Omega$ being the Rabi frequency of the target ion, thus quantifies the severity of crosstalk errors. As shown in Fig. S4, we numerically simulate the state fidelities of the original and error-mitigated ${\rm MS}_{YY}$ and ${\rm MS}_{ZZ}$ gates. As expected, the numerical results show that ${\rm MS}_{YY}$ gates, either original or error-mitigated ones, are insensitive to the crosstalk errors, while the fidelities of ${\rm MS}_{ZZ}$ gates degrade as the severity of crosstalk errors increases. According to the numerical results, the crosstalk residual error rate is about $0.68\times10^{-3}$ at the experimental level of qubit crosstalk. Finally, the remaining part of the residual error rate, $0.08\times 10^{-3}$, is attributed to the time-correlated systematic drift.






\end{document}